# A Novel Multi-Secret Sharing Approach for Secure Data Warehousing and On-Line Analysis Processing in the Cloud

Varunya Attasena, Nouria Harbi and Jérôme Darmont
Université de Lyon (Laboratoire ERIC), France

## ABSTRACT

Cloud computing helps reduce costs, increase business agility and deploy solutions with a high return on investment for many types of applications, including data warehouses and on-line analytical processing. However, storing and transferring sensitive data into the cloud raises legitimate security concerns. In this paper, we propose a new multi-secret sharing approach for deploying data warehouses in the cloud and allowing on-line analysis processing, while enforcing data privacy, integrity and availability. We first validate the relevance of our approach theoretically and then experimentally with both a simple random dataset and the Star Schema Benchmark. We also demonstrate its superiority to related methods.

## INTRODUCTION

Business intelligence (BI) has been an ever-growing trend for more than twenty years, but the recent advent of cloud computing now allows deploying data analytics even more easily. While building a traditional BI system typically necessitates an important initial investment, with the cloud pay-as-you-go model, users can punctually devote small amounts of resources in return for a one-time advantage. This trend is currently supported by numerous “BI as a service” offerings, with high economic stakes.

Although cloud computing is currently booming, data security remains a top concern for cloud users and would-be users. Some security issues are inherited from classical distributed architectures, e.g., authentication, network attacks and vulnerability exploitation, but some directly relate to the new framework of the cloud, e.g., cloud service provider or subcontractor espionage, cost-effective defense of availability and uncontrolled mashups (Chow et al., 2009). In the context of cloud BI, privacy is of critical importance. Security issues are currently handled by cloud service providers (CSPs). But with the multiplication of CSPs and subcontractors in many countries, intricate legal issues arise, as well as another fundamental issue: *trust*. Telling whether trust should be placed in CSPs falls back onto end-users, with the implied costs.

Critical security concerns in (especially public) cloud storage are depicted in Figure 1. User data might be deleted, lost or damaged. First, some CSPs have the policy of taking the highest profit. Therefore, unmodified or unaccessed data may be deleted to serve other customers. Second, data loss may also be caused by accidental, e.g., electrical or network failure, or intentional plans, e.g., maintenance or system backup. Moreover, virtual cloud architectures might not be sufficiently safeguarded from inside attacks. Finally, all CSPs cannot guarantee 100% data availability, although some cloud businesses must run on a 7/24 basis. Thus, data privacy, availability and integrity are major issues in cloud data security.

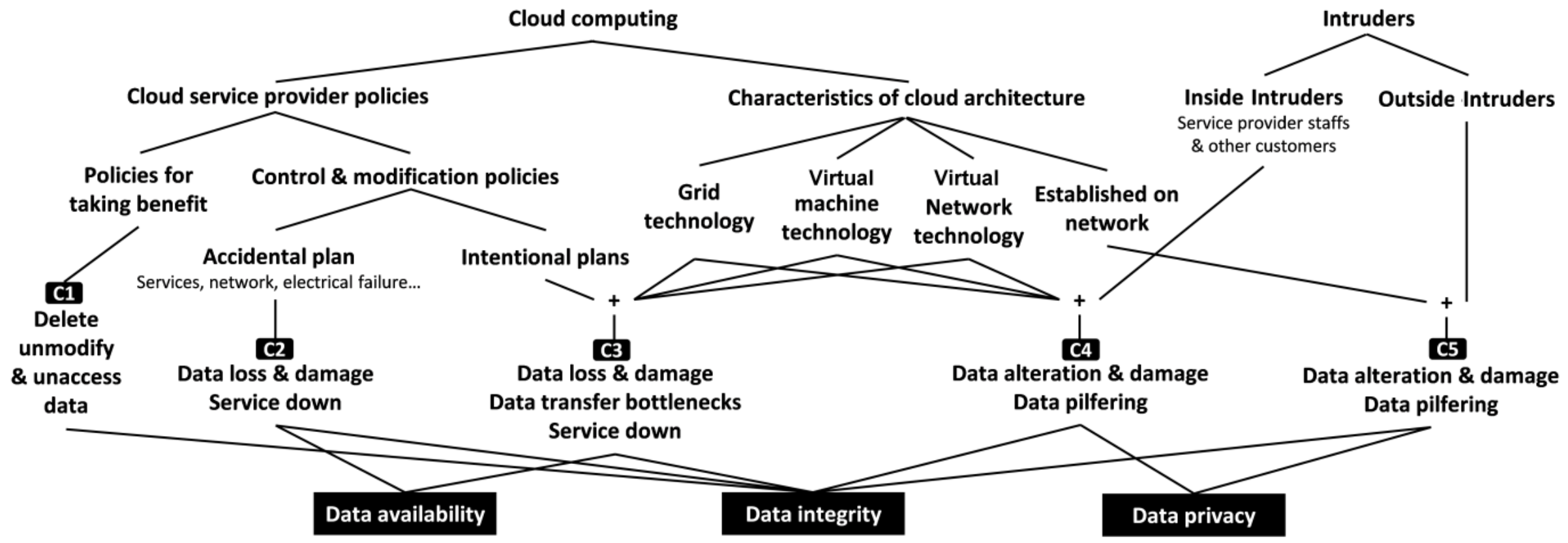

*Figure 1.* Cloud data security issues.

Encrypting and replicating data can solve most of these issues, but existing solutions are greedy in resources such as data storage, memory, CPU and bandwidth. Moreover, cloud data warehouses (DWs) must be both highly protected and effectively refreshed and analyzed through on-line analysis processing (OLAP). Thence, while CSPs must optimize service quality and profit, users seek to reduce storage and access costs within the pay-as-you-go paradigm. Thus, in cloud DWs, the tradeoff between data security and large-scale OLAP analysis poses a great challenge (Chow et al., 2009; Sion, 2007).

To address this challenge, we propose a global approach that relies on a new multi-secret sharing scheme, a family of encryption methods that enforce privacy and availability by design. Moreover, we incorporate in our approach features for data integrity verification and computation on shared data (or shares). Eventually, we minimize shared data volume. This paper expands (Attasena et al., 2013) along three axes. First, we complement the state of the art and deepen our analysis of related works. Second, we detail the section related to sharing a DW and specify the way OLAP queries run on shares. Finally, we complement our validation effort with new experiments, especially with the Star Schema Benchmark.

The remainder of this paper is organized as follows. We first introduce and discuss previous research related to our proposal. Based on this diagnosis, we further motivate and position our work. Then, we detail our secret sharing-based approach, before providing a security analysis and performance evaluation that highlight the relevance of our proposal and demonstrates the enhancements it brings over existing methods. We finally conclude this paper and hint at future research perspectives.

## RELATED WORKS

Existing research solve data privacy, availability and integrity issues by encrypting, anonymizing, replicating or verifying data (Figure 2).

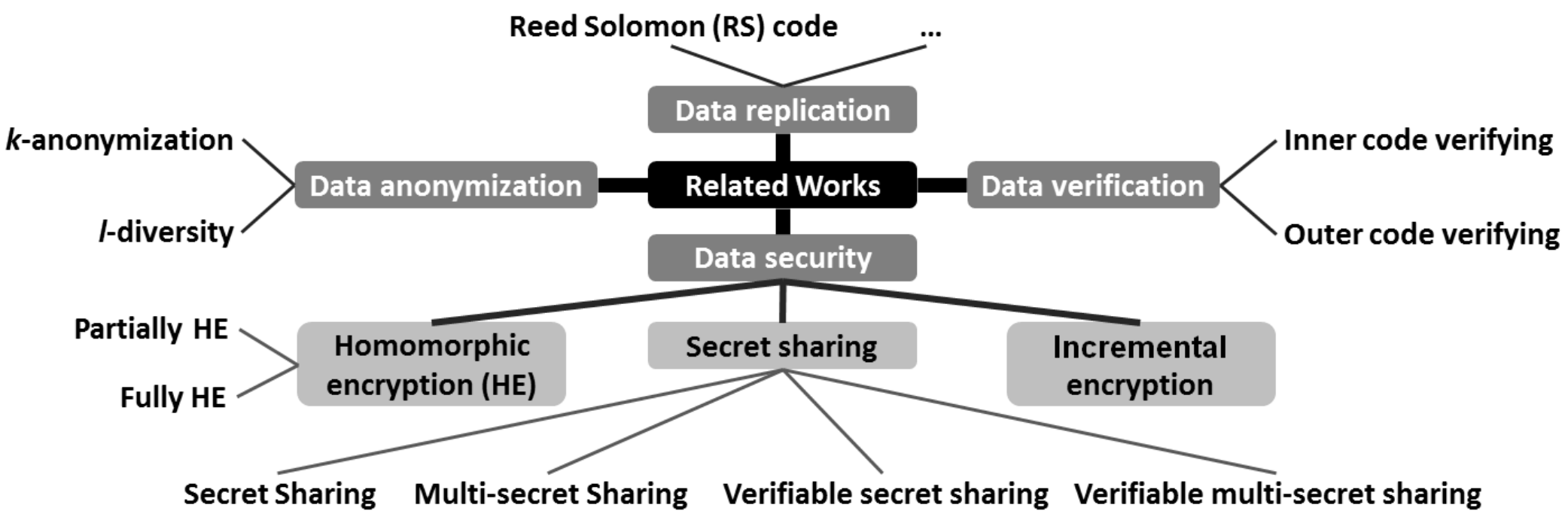

*Figure 2.* Existing data security solutions.

Encryption turns original data into unreadable cipher-text. Modern encryption schemes, such as homomorphic (HE – Melchor et al., 2008; Gentry, 2009) and incremental encryption (Bellare et al., 1994), help perform computations and updates on cipher-texts without decrypting them first. Partially HE allows only one operation, e.g., addition or multiplication, whereas fully HE supports several, but still does not allow mixed-operators. Unfortunately, HE is currently too computationally expensive for practical use. An older, well-known encryption strategy is secret sharing (Asmuth & Bloom, 1983; Blakley, 1979; Shamir, 1979), which distributes individually meaningless shares of data to $n$ participants to enforce privacy. A subset of $t \leq n$ participants is required to reconstruct the secret. Moreover, up to $n - t$ participants may disappear without compromising data availability. The drawback of this solution is the multiplication of the initial data volume by the number of participants. Modern secret sharing schemes, such as multi-secret sharing (Liuet al., 2012; Waseda & Soshi, 2012), verifiable secret sharing (Bu & Zhou, 2009), and verifiable multi-secret sharing (Bu & Yang, 2013; Eslami & Ahmadabadi, 2010; Hu et al., 2012), help reduce the volume of shares, verify the honesty of each participant, and both, respectively.

Data anonymization (Cormode & Srivastava, 2009; Kenneally & Claffy, 2010; Machanavajjhala et al., 2007; Sweeney, 2002) is also used to enforce data privacy. In a database, only keys or sensitive information are protected (Sedeyao, 2012). Thus, data anonymization straightforwardly allows data querying. There are several models (e.g., $k$-anonymized, $l$-diversity) and techniques (hashing, hiding, permutation, shift…) to protect keys and sensitive information, respectively. For example, the $k$-anonymized model transforms $k$ distinguishable records into $k$ indistinguishable records (Sweeney, 2002). The $l$-diversity model creates $l$ different sensitive values from only one value in each key identification combination (Machanavajjhala et al., 2007). While cheap when accessing data, anonymization is not strong enough to protect against attacks such as homogeneity and background knowledge attacks (Sedeyao, 2012), and is not designed to address data availability and integrity issues.

Data replication (Padmanabhan et al., 2008) is the process of copying some or all data from one location to one or several others. Its main purposes are to improve availability, fault-tolerance and/or accessibility. A well-known data replication scheme is Reed Solomon (RS) code (Thomas & Schwarz, 2002), which is quite similar to secret sharing. RS code indeed distributes data amongst a group of participants and can reconstruct data even if some participants disappear, thus enforcing availability. RS code and secret sharing mostly differ in their driving goals, i.e., availability and privacy, respectively.

Data verification (Bowers et al., 2009; Juels & Kaliski, 2007; Shacham & Waters, 2008; Wang et al., 2009) is the process of checking data integrity, by verifying data corruption caused by either accident or intruder attack, with the help of signatures (digital signature, message authentication, fingerprint…). However, since signature creation typically involves random or hash functions, they cannot guarantee 100% data correctness. Moreover, so-called outer code verifying methods (Juels & Kaliski, 2007) allow checking encrypted data without decrypting them first.

Table 1 summarizes the features of the above security approaches, with respect to data privacy, availability, integrity and full access. No existing approach simultaneously satisfies all criteria.

| Approaches | Data Privacy | Data availability | Data Integrity | Data access |
|---|---|---|---|---|
| **Encryption** | | | | |
| - Homomorphic encryption | √ | | | On encrypted data without decryption. |
| - Incremental encryption | √ | | | On encrypted data without decryption. |
| - Secret sharing | √ | √ | | Summing and averaging shares |
| - Multi-secret sharing | √ | √ | | |
| - Verifiable secret sharing | √ | √ | √ | |
| - Verifiable multi-secret sharing | √ | √ | √ | |
| **Data anonymization** | √ | | | On non-anonymized data |
| **Data replication**(RS code) | | √ | | |
| **Data verification** | | | √ | |

*Table 1.* Comparison of data security solutions.

Eventually, some security solutions directly relate to ours. Most apply Shamir's (1979) classical secret sharing to relational databases or data warehouses (Emekci et al., 2006; Hadavi & Jalili, 2010), thus, enforcing data privacy, availability and updating. In addition, Thompson et al. (2009), Wang et al. (2011) and Hadavi et al. (2012) also support data verification through HE, a hash function, and checksums and a hash function, respectively. Most of these methods allow computing at least one query type (aggregation, range and match queries) on shares.

As in the three last cited approaches (here after denoted TWH for brevity), our strategy is to extend one security scheme presenting interesting characteristics, namely multi-secret sharing, by integrating the missing features needed in cloud DWs. However, in our approach, shared data volume is better controlled than in TWH's, i.e., it is significantly lower than $n$ times that of the original data volume. Moreover, we also incorporate both inner and outer code data verification in our solution, whereas TWH only feature inner code data verification. Finally, we also include capabilities from homomorphic and incremental encryption that allow updating and computing basic operations on shares. Thus, to the best of our knowledge, our multi-secret sharing-based approach is the first attempt at securing data warehousing and OLAP while minimizing data volume.

## MULTI-SECRET SHARING OF CLOUD DATA WAREHOUSES

The solution we propose is based on trusting neither CSPs nor network data transfers. It is subdivided into two schemes. Scheme-I is a new multi secret sharing scheme that transforms data into blocks (to optimize computing and storage costs), and shares data blocks at several CSPs'. Each CSP only stores part of the shares, which are not exploitable, neither by the CSP nor any intruder, because they have been transformed by a mathematical function. Though performing computations on shares is possible, i.e., data need not be decrypted, it yields meaningless results. It is only when all results are mathematically transformed back at the user's that they can be reconstructed into global, meaningful information. Individual shares and computed results being encrypted, network transfers to and from CSPs are thus safe. Hence, privacy is achieved at any point outside of the user's (network, providers). Finally, to verify the honesty of CSPs and the correctness of shares, we incorporate into Scheme-I two types of hash-based signatures. Signatures help verify data correctness in case some CSPs are not honest, and incorrect or erroneous data before decryption.

However, updating and querying data are still difficult and expensive in Scheme-I, because data pieces are dependent on the others in the same block. Thus, Scheme-II builds upon Scheme-I to actually allow sharing and querying a DW in the cloud. Assuming a DW stored in a relational database, each attribute value in each record is shared independently. We first transform each attribute value to at least one block, depending on data type and size (e.g., one block for integers, reals or characters; and $l$ blocks for strings of length $l$), and encrypt each data block with Scheme-I. Then, we allow analyzing data over shares with ad-hoc queries and Relational OLAP (ROLAP) operations, without decrypting all data first whenever possible. All basic OLAP operations (roll-up, drill-down, some slice and dice, pivot and drill-across) can apply directly on shares at the CSPs', with results being reconstructed at the user's. However, other complex queries must be transformed or split first, depending on operations and functions used.

### Scheme-I: $(m, n, t)$ Multi-secret sharing with data verification

Scheme-I is an $(m, n, t)$ multi-secret sharing scheme: $m$ data pieces are encrypted and shared among $n$ CSPs. $t$ out of $n$ shares can reconstruct the original data. The total volume of shares is only about $mn/(t-1)$. Data are organized into blocks that are encrypted and decrypted all at once. The priorities of blocks and data in the blocks are important because they directly affect the results of data access in Scheme-II. All data pieces in a block are encrypted at once by $n$ distinct random $t$-variable linear equations, where variables are data and their signatures and coefficients are pseudorandom. Eventually, we introduce two types of signatures. The first, inner signature is created from all data pieces in one block. It matches with data in the reconstruction process if CSPs return correct shares. The second, outer signature is created from each share. At each CSP's, it verifies shares before transferring them back to the user for reconstruction.

Parameters of Scheme-I are listed in Table 2. $ID_{i=1..m}$ are randomly selected from distinct integers and are stored at the user's. $D$ is split into $o$ blocks with $o = \left\lceil \frac{m}{t-1} \right\rceil$. If $m$ is not a multiple of $t-1$, the last block is padded with integer values -1 (Figure 3).

| Parameters | Definitions |
|---|---|
| $n$ | Number of CSPs |
| $CSP_k$ | CSP number $k$ |
| $m$ | Number of data pieces |
| $o$ | Number of data blocks |
| $t$ | Number of shares necessary for reconstructing original data |
| $P$ | A big prime number |
| $D$ | Original data such that $D = \{d_1, \ldots, d_m\}$ and $D = \{b_1, \ldots, b_o\}$ |
| $d_i$ | The $i^{th}$ piece of $D$ in integer format such that $P - 2 > d_i \geq 0$ |
| $b_j$ | The $j^{th}$ block of $D$ such that $b_j = \{d_{(j-1)(t-1)}, \ldots, d_{(j)(t-1)}\}$ |
| $ID_k$ | Identifier number of $CSP_k$ such that $ID_k > 0$ |
| $e_{j,k}$ | Share of $b_j$ stored at $CSP_k$ |
| $s_in_j$ | Signature of original data in $b_j$ such that $P > s_in_j \geq 0$ |
| $s_out_{j,k}$ | Signature of share of $b_j$ stored at $CSP_k$ |

*Table 2.* Scheme-I parameters.

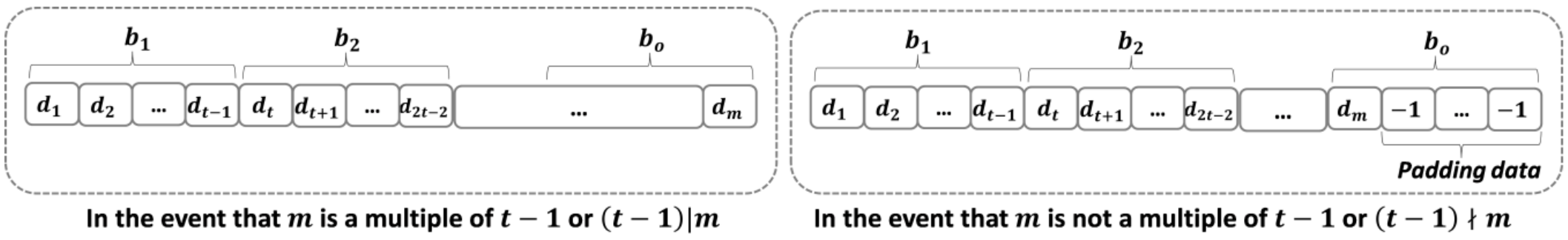

*Figure 3.* Organization of data in blocks.

***Data sharing process***

Each data block is encrypted independently (Figure 4). Data pieces in block $b_j$ are encrypted as follows.

1. Compute signature $s_in_j$ from data block $b_j$ with homomorphic function $H_1$: $s_in_j = H_1(b_j)$.
2. Create $n$ distinct random $t-1$ linear equations (Equation 1).
   $y = f_k(x_1, \cdots, x_t) = \left(\sum_{h=1}^{t-1}(x_h + 2) \times a_{k,h}\right) + (x_t \times a_{k,t})$ (1)
   where $x_i$ is positive variable, $a_{kh}$ is the $h^{\text{th}}$ positive pseudorandom coefficient seeded at $ID_k$, $P > a_{k,h} \geq 0$ and $f_{k1} \neq f_{k2}$ if $k1 \neq k2$. These functions are used for all blocks.
3. Compute the set of shares $\{e_{j,k}\}_{k=1\ldots n}$ from data block $b_j$ such that $e_{j,k} = f_k(b_j, s_in_j)$, and distribute each share $e_{j,k}$ to $CSP_k$.
4. Compute signatures $\{s_out_{j,k}\}_{k=1\ldots n}$ with hash function $H_2$ such that $s_out_{j,k} = H_2(e_{j,k})$, and distribute each signature $s_out_{j,k}$ to $CSP_k$ along with $e_{j,k}$.

Thus, data and their signatures are shared among $n$ CSPs. $CSP_k$ stores $o$ pairs of shares and signatures $\left((e_{j,k}, s_out_{j,k})_{j=1\ldots o}\right)$.

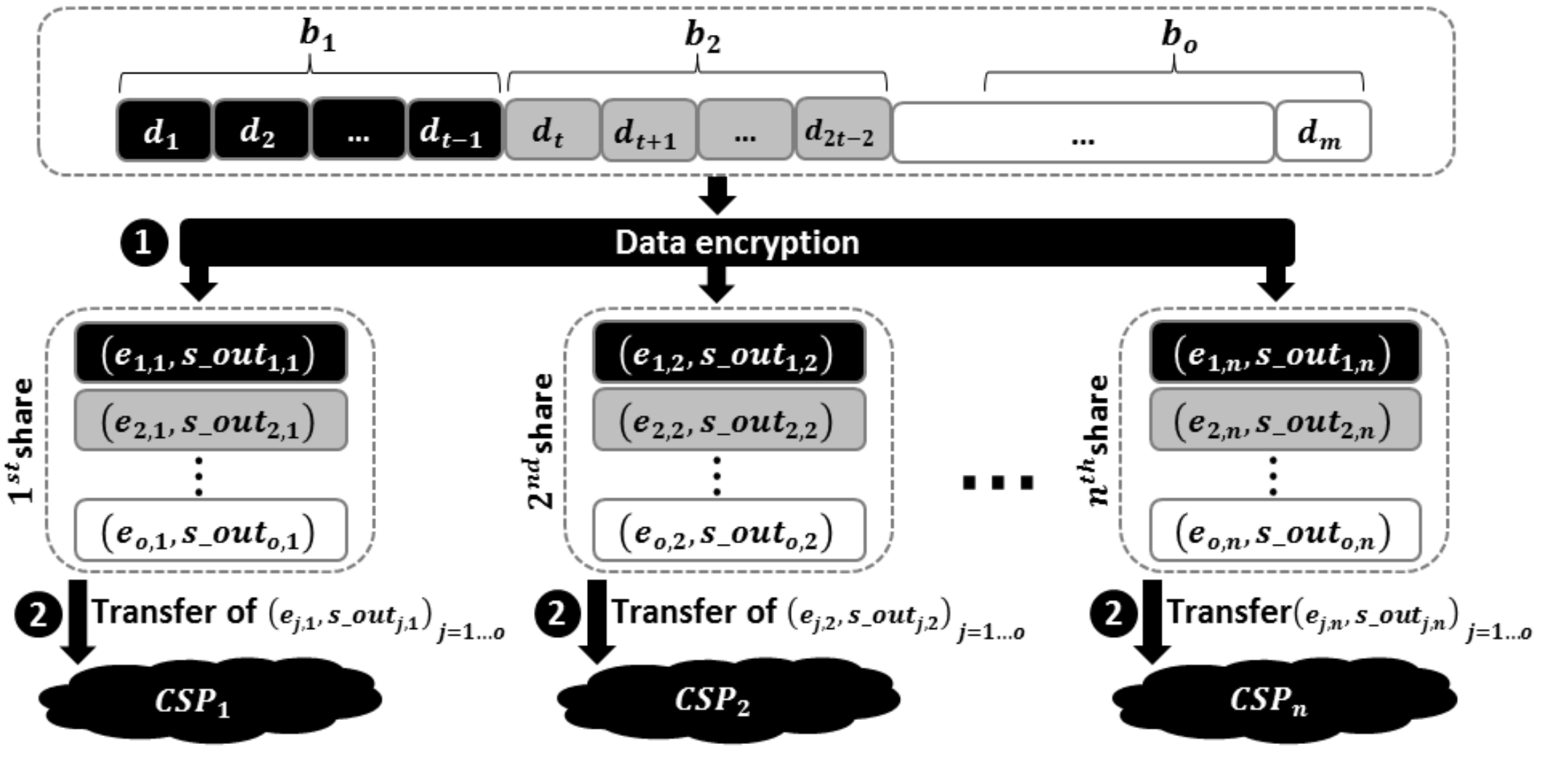

*Figure 4.*Data sharing process.

***Data reconstruction process***

A dataset $D$ is reconstructed from shares and signatures $\left(e_{j,k}, s_out_{j,k}\right)_{j=1\ldots o}$ stored at $CSP_k \in G$, where $G$ is any group of $t$ CSPs (Figure 5). There are two phases to reconstruct original data: the initialization phase and the actual reconstruction phase.

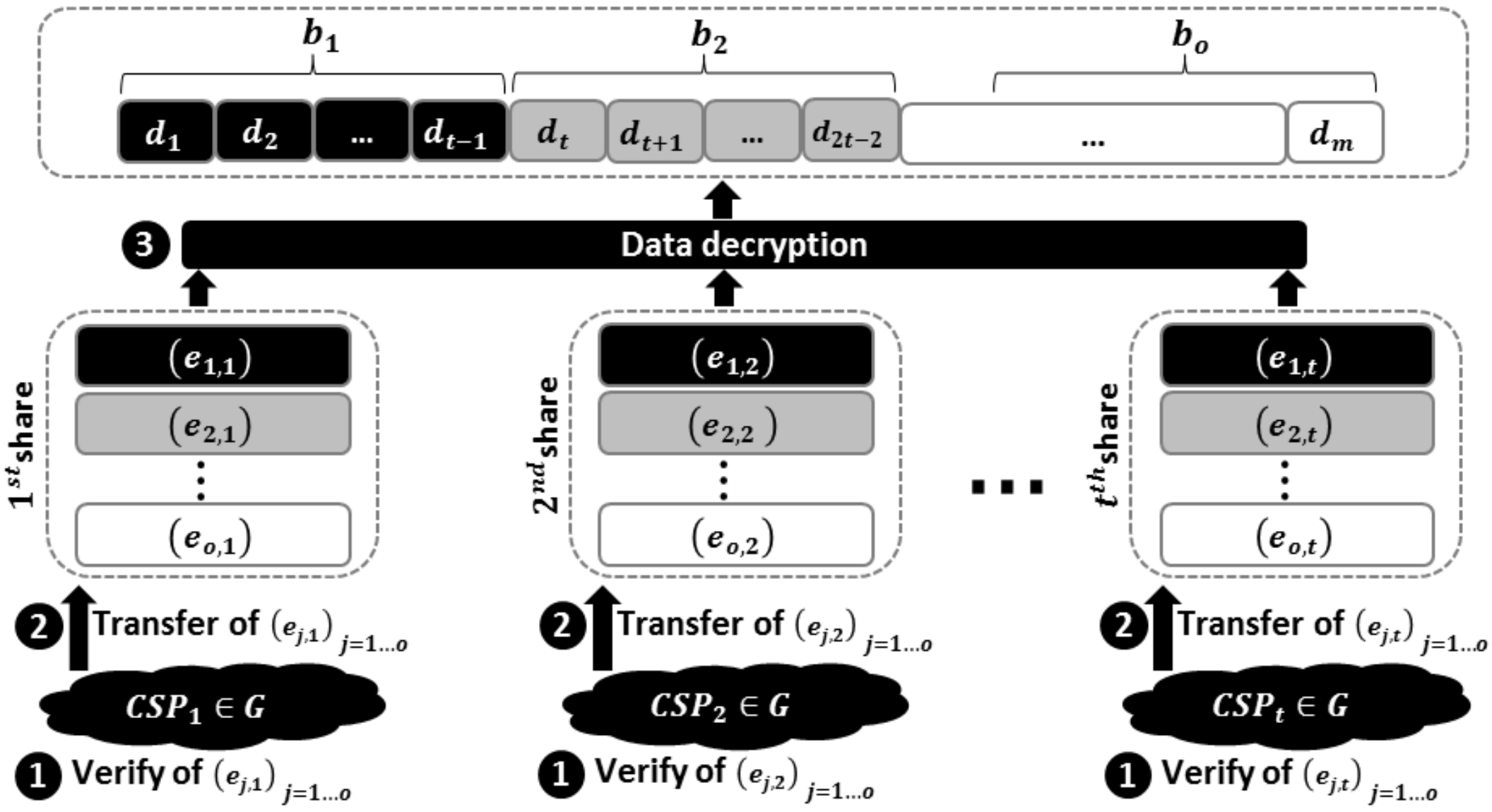

*Figure 5.*Data reconstruction process.

*Initialization phase:* In this phase, share correctness is verified and a matrix $C$ that is used in the reconstruction phase is created as follows.

1. Verify information at all $CSP_k \in G$. At each CSP's, only shares to be decrypted are verified for correctness. Share $e_{j,k}$ is correct if $s_out_{j,k} = H_2\left(e_{j,k}\right)$. In case of error at $CSP_k$, then another CSP is selected and correctness is verified again.
2. At the user's, matrix $A$ is created from $ID_k$ of $CSP_k \in G$ such that $A = \left[a_{x,y}\right]_{t\times t}$, where $a_{x,y}$ is the $y^{th}$ coefficient of $f_x$. Then, $C$ is computed such that $C = A^{-1}$. Let $c_{x,y}$ be an entry in the $x^{th}$ row and the $y^{th}$ column of matrix $C$.

*Reconstruction phase:* To decrypt data block$b_j$, share $e_{j,k}$ of $CSP_k \in G$ is transferred to the user and decrypted as follows.

1. Compute data block $b_j$ (Equation 2) and its signature $s_in_j$ (Equation 3).
   $$d_{(j-1)(t-1)+l} = \left(\sum_{h=1}^{t} c_{l,h} \times e_{j,h}\right) - 2; \forall[1, t-1] \quad (2)$$
   $$s_in_j = \sum_{h=1}^{t} c_{t,h} \times e_{j,h} \quad (3)$$
2. If $s_in_j = H_1\left(b_j\right)$, then data in block $b_j$ are correct. In case of errors, the user can re-construct data from shares from a new $G$.

**Scheme-II: Sharing a data warehouse in the cloud**

In this section, we exploit Scheme-I to share a DW among CSPs. Databases attribute values, except NULL values and primary or foreign keys, are shared in relational databases at CSPs'. Keys help match records in the data reconstruction process and perform JOIN and GROUP BY operations. Any sensitive primary key, such as a social security number, is replaced by an unencrypted sequential integer key. Each attribute in the original tables is transformed into two attributes in encrypted tables, i.e., share and signature attributes. Figure 6 shows the example of a PRODUCT table that is shared among three CSPs. To handle data from a shared DW, we propose solutions to encrypt data of various types, to share the customary DW logical models in the cloud, to perform loading, backup and recovery processes, and to analyze shared data through ROLAP operations.

| ProdNo | ProName | ProdDescr | CategoryID | UnitPrice |
|---|---|---|---|---|
| 124 | Shirt | Red | 1 | 75 |
| 125 | Shoe | *NULL* | 2 | 80 |

(a) Original data

| ProdNo | ProName | SigPN | ProdDescr | SigPD | CategoryID | UnitPrice | SigUP |
|---|---|---|---|---|---|---|---|
| 124 | {29,18,21,22,28} | {1,4,0,1,0} | {26,20,17} | {5,6,3} | 1 | 16 | 2 |
| 125 | {29,18,13,20} | {1,4,6,6} | *NULL* | *NULL* | 2 | 20 | 6 |

(b) Shares at $CSP_1$

| ProdNo | ProName | SigPN | ProdDescr | SigPD | CategoryID | UnitPrice | SigUP |
|---|---|---|---|---|---|---|---|
| 124 | {29,16,19,46,52} | {1,2,5,4,3} | {26,45,42} | {5,3,0} | 1 | 43 | 1 |
| 125 | {29,16,37,45} | {1,2,2,3} | *NULL* | *NULL* | 2 | 20 | 6 |

(c) Shares at $CSP_2$

| ProdNo | ProName | SigPN | ProdDescr | SigPD | CategoryID | UnitPrice | SigUP |
|---|---|---|---|---|---|---|---|
| 124 | {33,22,25,39,45} | {5,1,4,4,3} | {30,37,34} | {2,2,6} | 1 | 33 | 5 |
| 125 | {33,22,30,37} | {5,1,2,2} | *NULL* | *NULL* | 2 | 24 | 3 |

(d) Shares at $CSP_3$

*Figure 6.* Example of original data and shares at three CSPs'.

***Data types***

To handle the usual data types featured in databases, we encrypt and handle each data piece independently. Data pieces of any type are first transformed into integers, then split into one or several data blocks (depending on type and value), and finally encrypted with Scheme-I.

For sharing an integer, date or timestamp $I$, $I$ is split into $t-1$ pieces $d_{i=1\ldots(t-1)}$ such that $d_i = \left\lfloor \frac{I}{p^{i-1}} \right\rfloor \bmod p$, where $p$ is a prime number and $\|p\| > \frac{\|maxint\|}{t-1}$ bits, where $\|maxint\|$ is the size of the maximum integer value in bits. Then, $d_{i=1\ldots(t-1)}$ is encrypted to $n$ shares with Scheme-I.

For sharing a real $R$, $R$ is transformed into an integer $I$ by multiplication. For example, let $R$ be stored in numeric format $(v,\ s)$, where $v$ is a precision value and $s$ a scale value. Then, $R$ is transformed into $I = R \times 10^{|s|}$. $I$ can then be encrypted as any integer.

For sharing a character $L$, $L$ is transformed into an integer $I$through its ASCII code. For example, let $L$ be 'A'. $L$ is transformed into $I = 65$. $I$ can then be encrypted as any integer.

For sharing a string $S$, $S$ is transformed into a set of integers $\{I_j\}_{j=1\ldots l}$ where $l$ is the length of $S$, using the ASCII code of each character in $S$. For example, let $S$ be 'ABC'. Then, $S$ is transformed into $\{I_j\}_{j=1\ldots 3} = \{65, 66, 67\}$. After transformation, each character $I_j$ is encrypted independently as any integer.

For sharing a binary string $B$, $B$ is transformed into a set of integers $\{I_j\}_{j=1\ldots\left\lceil \frac{l}{\|maxint\|} \right\rceil}$ where $l$ is the length of $B$ and $\|maxint\|$ is the size of the maximum integer value in bits. For example, let $B = 1000000000000000000000000000000011$ and $\|maxint\| = 32$ bits. Then $B$ is split into two smaller binaries: 10 and 00000000000000000000000000000011 sizing less

than $\|maxint\|$, which are then transformed into $\{I_j\}_{j=1,2} = \{2, 3\}$. After transformation, each $I_j$ is encrypted independently as any integer.

An example of sharing an integer $I$ follows.

1. Sharing parameters are assigned as follows: $n = 4, t = 3$ and $p = 13$.
2. Homomorphic and hash functions are $H_1(b_j) = \sum_{d_i \in b_j} d_i (\mathrm{mod}\, p)$ and $H_2(e_{j,k}) = e_{j,k} \bmod 13$.
3. Let $I = 75$, i.e., the shirt's unit price in Figure 6(a).
4. We compute $d_{i=1,2}$ as follows: $d_1 = \left\lfloor \frac{75}{13^{1-1}} \right\rfloor \bmod 13 = 10$ and $d_2 = \left\lfloor \frac{75}{13^{2-1}} \right\rfloor \bmod 13 = 5$.
5. We compute $s_in_1 = H_1(b_1) = (10 + 5) \bmod 13 = 2$.
6. Let four random 3-variable linear equations be:
   a. $y = f_1(x_1, x_2, x_3) = 1 \times (x_1 + 2) + 0 \times (x_2 + 2) + 2 \times x_3$,
   b. $y = f_2(x_1, x_2, x_3) = 3 \times (x_1 + 2) + 1 \times (x_2 + 2) + 0 \times x_3$,
   c. $y = f_3(x_1, x_2, x_3) = 2 \times (x_1 + 2) + 1 \times (x_2 + 2) + 1 \times x_3$,
   d. $y = f_4(x_1, x_2, x_3) = 0 \times (x_1 + 2) + 2 \times (x_2 + 2) + 1 \times x_3$.
7. We compute $e_{1,k=1\ldots4}$ such that $e_{1,1} = f_1(10,5,2) = 1 \times (10 + 2) + 0 \times (5 + 2) + 2 \times 2 = 16$. Similarly, $e_{1,2} = 43$, $e_{1,3} = 33$ and $e_{1,4} = 14$.
8. We compute $s_out_{1,k=1\ldots4}$ such that $s_out_{1,1} = H_2(16) = 16 \bmod 7 = 2$. Similarly, $s_out_{1,2} = 1$, $s_out_{1,3} = 5$ and $s_out_{1,4} = 0$.
9. We distribute each couple $(e_{1,k}, s_out_{1,k})$ to $CSP_k$.

Then, $I$ is reconstructed as follows.

1. Suppose $CSP_1$, $CSP_2$ and $CSP_3$ are selected into $G$.
2. We verify $s_out_{1,j=1,2,3}$ such that $s_out'_{1,1} = H_2(16) = 16 \bmod 7 = 2 = s_out_{1,1}$. Then $e_{1,1}$ is correct. After verification, all three shares $\{e_{1,1}, e_{1,2}, e_{1,3}\}$ are found correct.
3. We create matrix $A$ from $ID_{i=1,2,3}$: $A = \begin{bmatrix} 1 & 0 & 2 \\ 3 & 1 & 0 \\ 2 & 1 & 1 \end{bmatrix}$.
4. We compute matrix $C = A^{-1} \bmod P = \dfrac{\begin{bmatrix} 1 & 2 & -2 \\ -3 & -3 & 6 \\ 1 & -1 & 1 \end{bmatrix}}{3}$
5. We compute $d_{i=1,2}$ as follows.
   a. $d_1 = ((16 \times 1 + 43 \times 2 + 33 \times -2)/3) - 2 = 10$.
   b. $d_2 = ((16 \times -3 + 43 \times -3 + 33 \times 6)/3) - 2 = 5$.
6. We compute $s_in_1 = (16 \times 1 + 43 \times -1 + 33 \times 1)/3 = 2$.
7. We verify the original data. The result is correct since $s_in'_1 = H_1(b_1) = (10 + 5) \bmod 13 = 2 = s_in_1$.

***Data warehouse sharing***

Since each table of a shared DW is stored in a relational database at a given CSP's and each attribute value in each record is encrypted independently, Scheme-II straightforwardly helps implement any DW logical model, i.e., star, snowflake or constellation schema. Figures 7(a) and 7(b) show an example of snowflake-modeled DW that is shared among three CSPs. Each shared DW bears the same schema as the original DW's, but type and size of each attribute in each shared table differ from the original tables. All attribute types, except Booleans that are not encrypted to save computation and data storage costs, are indeed transformed into integers.

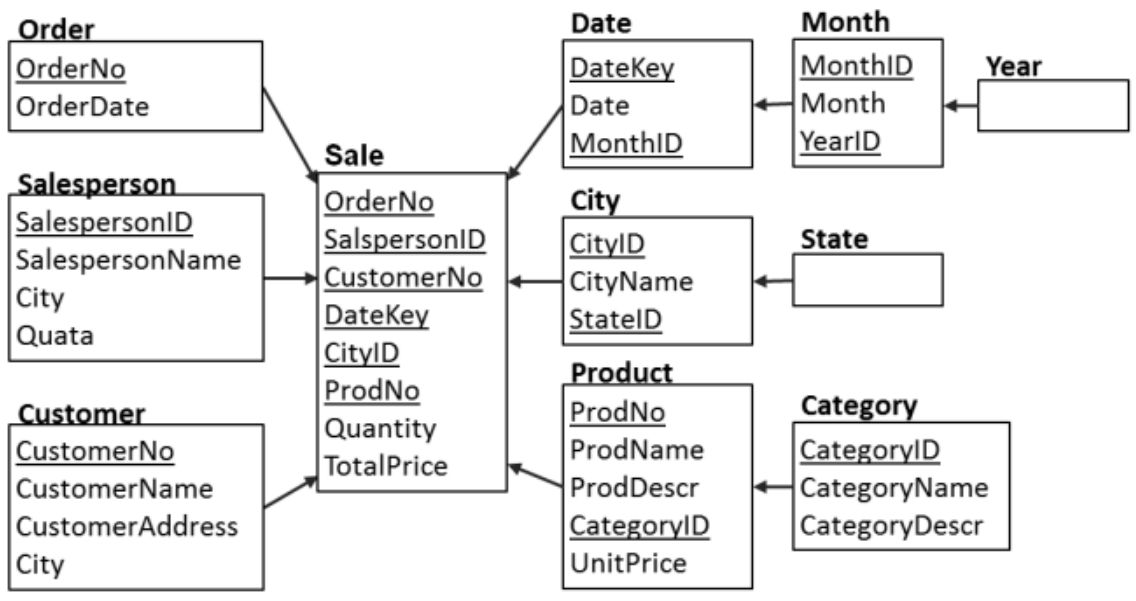

(a) Original schema

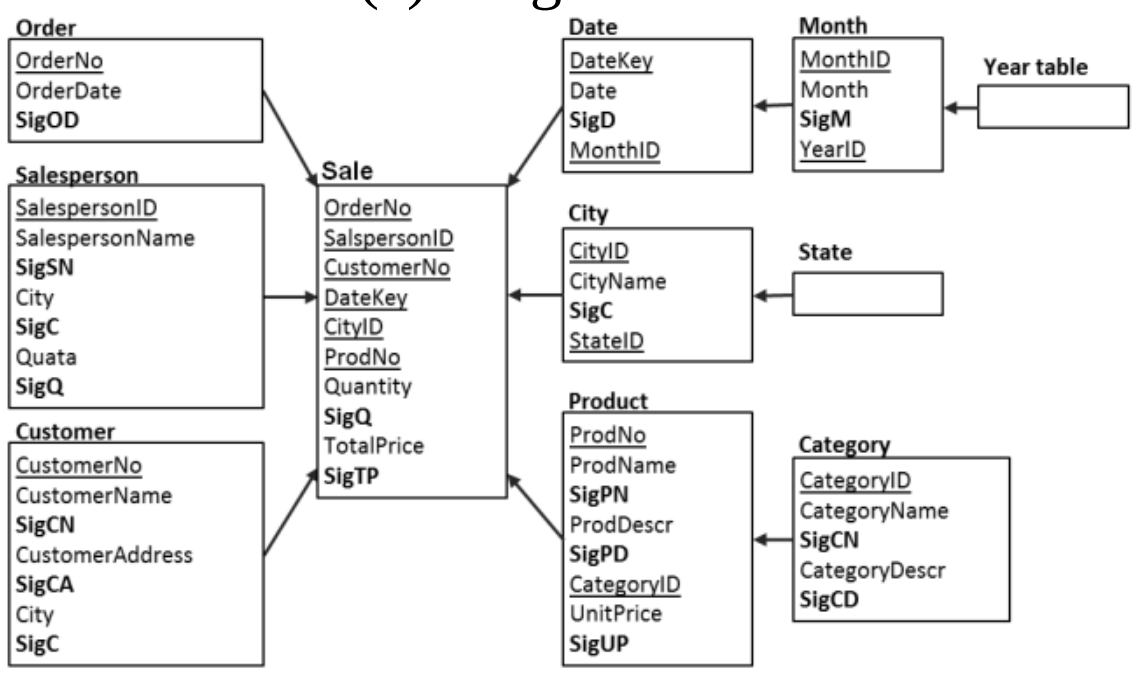

(b) Transformed schemas at CSPs'

| foreign keys | | | | | Shares and Signatures | | | |
|---|---|---|---|---|---|---|---|---|
| Time Attributes | | | Product Attributes | | Aggregation Attributes | | Signature Attributes | |
| YearID | MonthID | DateID | CategoryID | ProdNo | TotalPrice | Number | SigTotalPrice | SigNumber |
| NULL | NULL | NULL | NULL | NULL | 432 | 582 | 6 | 3 |
| NULL | NULL | NULL | 1 | NULL | 486 | 124 | 4 | 2 |
| NULL | NULL | NULL | 1 | 1 | 763 | 632 | 6 | 6 |
| NULL | NULL | NULL | 1 | ⋮ | ⋮ | ⋮ | ⋮ | ⋮ |
| NULL | NULL | NULL | 2 | 1 | 243 | 243 | 3 | 8 |
| NULL | NULL | NULL | ⋮ | ⋮ | ⋮ | ⋮ | ⋮ | ⋮ |
| 1 | NULL | NULL | NULL | NULL | 423 | 234 | 8 | |
| 1 | NULL | NULL | 1 | NULL | 124 | 211 | 5 | 3 |
| 1 | NULL | NULL | 1 | 1 | 342 | 123 | 3 | 9 |
| 1 | NULL | NULL | 1 | ⋮ | ⋮ | ⋮ | ⋮ | ⋮ |
| 1 | NULL | NULL | 2 | 1 | 234 | 832 | 9 | 2 |
| 1 | NULL | NULL | ⋮ | ⋮ | ⋮ | ⋮ | ⋮ | ⋮ |
| 1 | 1 | NULL | NULL | NULL | 962 | 248 | 2 | 9 |
| 1 | 1 | NULL | 1 | NULL | 244 | 724 | 3 | 0 |
| 1 | 1 | NULL | 1 | 1 | 164 | 257 | 7 | 5 |
| 1 | 1 | NULL | 1 | ⋮ | ⋮ | ⋮ | ⋮ | ⋮ |
| 1 | 1 | NULL | 2 | 1 | 834 | 243 | 1 | 3 |
| 1 | 1 | NULL | ⋮ | ⋮ | ⋮ | ⋮ | ⋮ | ⋮ |
| 1 | 2 | NULL | NULL | NULL | 174 | 262 | 9 | 4 |
| 1 | ⋮ | NULL | ⋮ | ⋮ | ⋮ | ⋮ | ⋮ | ⋮ |
| 1 | 1 | 1 | NULL | NULL | 335 | 353 | 0 | 2 |
| 1 | 1 | 1 | 1 | NULL | 266 | 742 | 3 | 7 |
| 1 | 1 | 1 | 1 | 1 | 145 | 262 | 2 | 8 |
| 1 | 1 | 1 | 1 | ⋮ | ⋮ | ⋮ | ⋮ | ⋮ |
| 1 | 1 | 1 | 2 | 1 | 268 | 458 | 1 | 4 |
| 1 | 1 | 1 | ⋮ | ⋮ | ⋮ | ⋮ | ⋮ | ⋮ |
| 1 | 1 | 2 | NULL | NULL | 292 | 534 | 8 | 9 |
| 1 | 1 | ⋮ | ⋮ | ⋮ | ⋮ | ⋮ | ⋮ | ⋮ |
| ⋮ | ⋮ | ⋮ | ⋮ | ⋮ | ⋮ | ⋮ | ⋮ | ⋮ |

(c) Sample shared cube-I

*Figure 7.* Example of shared data warehouse and cube.

Moreover, Scheme-II supports the storage of data cubes that optimize response time and bandwidth when performing ROLAP operations. Cubes are physically stored into tables that are shared among CSPs, retaining the same structure. For example, Figure 7(c) features a shared cube named cube-I that totalizes total prices and numbers of sales by time period and by product. Shared cubes include signatures for shared aggregate measures and customarily use NULL values to encode superaggregates. Finally, indices can also be shared to improve query performance. However, they must be created from the original data before the sharing process. We envisage lazy index creation on shares in future research, though.

***Loading, backup and recovery processes***

For loading data into a shared DW, each data piece is encrypted and loaded independently. New data can be loaded without decrypting previous data first, because each attribute value in each record is encrypted independently. For instance, in Figure 8, data from Figure 6 are already shared and the last record (#126) is new.

| ProdNo | ProName | ProdDescr | CategoryID | UnitPrice |
|---|---|---|---|---|
| 124 | Shirt | Red | 1 | 75 |
| 125 | Shoe | *NULL* | 2 | 80 |
| 126 | Ring | *NULL* | 1 | 80 |

(a) Original data

| ProdNo | ProName | SigPN | ProdDescr | SigPD | CategoryID | UnitPrice | SigUP |
|---|---|---|---|---|---|---|---|
| 124 | {29,18,21,22,28} | {1,4,0,1,0} | {26,20,17} | {5,6,3} | 1 | 16 | 2 |
| 125 | {29,18,13,20} | {1,4,6,6} | *NULL* | *NULL* | 2 | 20 | 6 |
| 126 | {26,21,10,26} | {5,0,3,5} | *NULL* | *NULL* | 1 | 20 | 6 |

(b) Shares at $CSP_1$

| ProdNo | ProName | SigPN | ProdDescr | SigPD | CategoryID | UnitPrice | SigUP |
|---|---|---|---|---|---|---|---|
| 124 | {29,16,19,46,52} | {1,2,5,4,3} | {26,45,42} | {5,3,0} | 1 | 43 | 1 |
| 125 | {29,16,37,45} | {1,2,2,3} | *NULL* | *NULL* | 2 | 20 | 6 |
| 126 | {26,19,34,51} | {5,5,6,2} | *NULL* | *NULL* | 1 | 20 | 6 |

(c) Shares at $CSP_2$

| ProdNo | ProName | SigPN | ProdDescr | SigPD | CategoryID | UnitPrice | SigUP |
|---|---|---|---|---|---|---|---|
| 124 | {33,22,25,39,45} | {5,1,4,4,3} | {30,37,34} | {2,2,6} | 1 | 33 | 5 |
| 125 | {33,22,30,37} | {5,1,2,2} | *NULL* | *NULL* | 2 | 24 | 3 |
| 126 | {30,25,27,43} | {2,4,6,1} | *NULL* | *NULL* | 1 | 24 | 3 |

(d) Shares at $CSP_3$

*Figure 8.* Example of sharing new data.

However, when updating cubes, some shared aggregates may have to be recomputed. Within Scheme-II, we currently cannot apply all aggregation operations on shares. Thus, such aggregations still require to be computed on the original data. For example, maximum and minimum cannot be computed on shares because original data order is lost in the sharing process. Averaging data must be performed by summing and counting. Hence, to optimize costs, aggregates are first computed on new data, and then aggregated to relevant existing shares, which are decrypted on-the-fly.

Finally, a backup process is unnecessary in our scheme, because each share $e_{j,l}$ is actually a backup share of all other shares $e_{j,k}$, where $k \in \{1, \ldots, l-1, l+1, \ldots, n\}$. In case a share is erroneous, it can be recovered from $t$ other shares.

***Data analysis over shares***

Since DWs and cubes can be shared in the cloud, Scheme-II directly supports all basic OLAP operations at the CSPs' through SQL operators and aggregation functions, and helps reconstruct the result on the user's side by performing queries on shared tables. For example, query "select YearID, YearName, TotalPrice from cube-I, year where cube-I.YearID=year.YearID and MonthID=null and DateID=null and CategoryID=null and ProdNo=null" can be run at *t* CSPs to compute the total price of products per year.

However, although some queries apply directly onto shares, others require some or all data to be decrypted. Simple SELECT/FROM queries directly apply onto shares. All join operators, when operating on unencrypted keys, also apply directly. However, when expressing conditions in a WHERE or HAVING clause, the following routine must be followed:

1. encrypt compared values,
2. substitute these shares to compared values in the query,
3. launch the query on *t* shares,
4. decrypt the *t* results,
5. reconstruct the global result by intersection.

For example, the query "SELECT ProdName FROM Product WHERE UnitPrice=75" would be transformed to "SELECT ProdName FROM Product WHERE UnitPrice=16" at $CSP_1$, where 16 is the share of 75 at $CSP_1$.

This routine works for many comparison operators (=, ≠, EXISTS, IN, LIKE…) and their conjunction, but when ordering is necessary, as in ORDER BY clauses and many comparison operators (>, <, ≥, ≤, BETWEEN…), it can no longer apply since the original order is broken when sharing data. Thus, all fetched data must be reconstructed at the client's before the result can be computed by an external program. However, some range queries can be transformed and performed on shares if comparison range is known and comparison attribute type is integer, char or string. For example, the query "SELECT ProdName FROM Product WHERE UnitPrice between 75 and 77" would be transformed to "SELECT ProdName FROM Product WHERE UnitPrice IN (16, 19, 22)" at $CSP_1$, where 16, 19 and 22 are the shares of 75, 76 and 77 at $CSP_1$., respectively.

Similarly, aggregation functions SUM, AVG and COUNT can directly apply on shares, whereas other aggregation functions, such as MAX and MIN, require all original data to be reconstructed prior to computation. Finally, grouping queries using the GROUP BY or GROUP BY CUBE clauses can directly apply if and only if they target unencrypted key attributes. Again, grouping by other attribute(s) requires all data to be reconstructed at the user's before aggregation by an external program.

Consequently, executing a complex query may require either transforming or splitting the query, depending on its clauses and operators, following the above guidelines. Figure 9 shows an example of complex query execution.

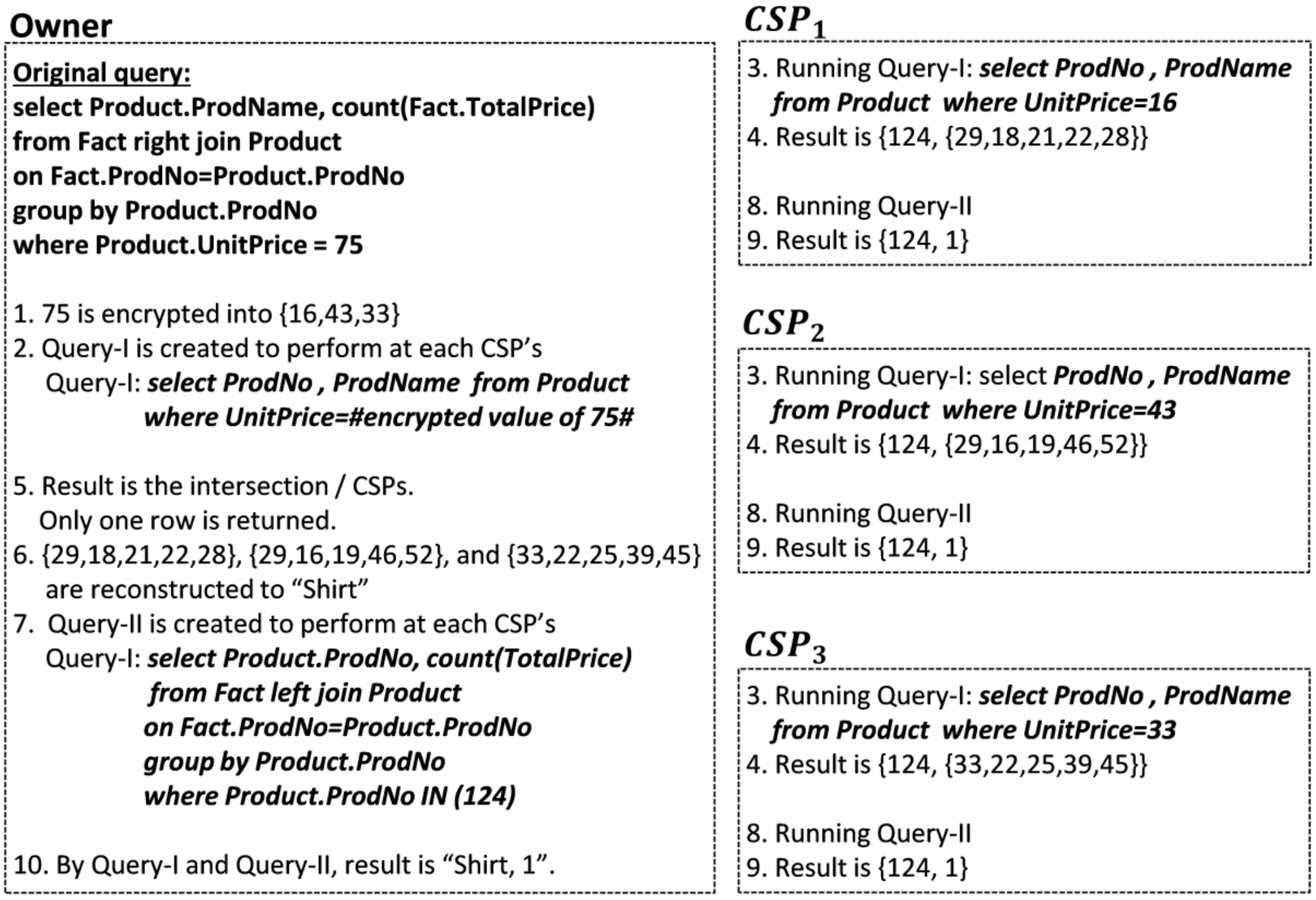

*Figure 9.* Example of complex query execution over shares.

## SECURITY ANALYSIS AND PERFORMANCE EVALUATION

In this section, we illustrate the relevance of our approach along two axes. First, we mainly theoretically study the security features of our schemes, which are our primary focus. Second, since our approach applies in the cloud, we both theoretically and experimentally study the factors that influence cost in the pay-as-you-go paradigm, i.e., computing, storage and data transfer costs, with respect to the TWH secret sharing schemes.

### Security analysis

#### *Privacy*

We focus here on data pilfering. Neither a CSP nor any intruder can decrypt original data from only one share, and data transferred between the user and CSPs are all encrypted. In case an intruder can steal shares from $x$ CSPs with $x \leq t$, the probability of discovering $b_j$ (the original data in the $j^{th}$ block) remains low, i.e., $\frac{1}{P^{2t-x-1}}$ and $\frac{1}{p^{2t-x-1}}$ in Scheme-I and Scheme-II, respectively (Figure 10). The probability of discovering $b_j$ depends on the following.

1. The size of control parameters $P$ in Scheme-I and $p$ in Scheme-II. In Scheme-I, the probability of breaking the secret is low because $P$ is a big prime number. In Scheme-II, the probability of breaking the secret ranges between $10^{-22}$ and $10^{-10}$ in Figure 10's example, because $p$ depends on $t$. If $\mathrm{p} = P$, the probabilities of breaking the secret are equal in both schemes, but storage cost in Scheme-II is not controlled, which falls back to Shamir's (1979) case.
2. The user-defined value of $t$. The higher $t$, the lower the probability of breaking the secret.
3. The number of pilfered shares $x$. The probability of breaking the secret obviously increases with $x$. However, both our schemes are secure enough since it is difficult to retrieve shares from at least $t$ CSPs by attacking them simultaneously.

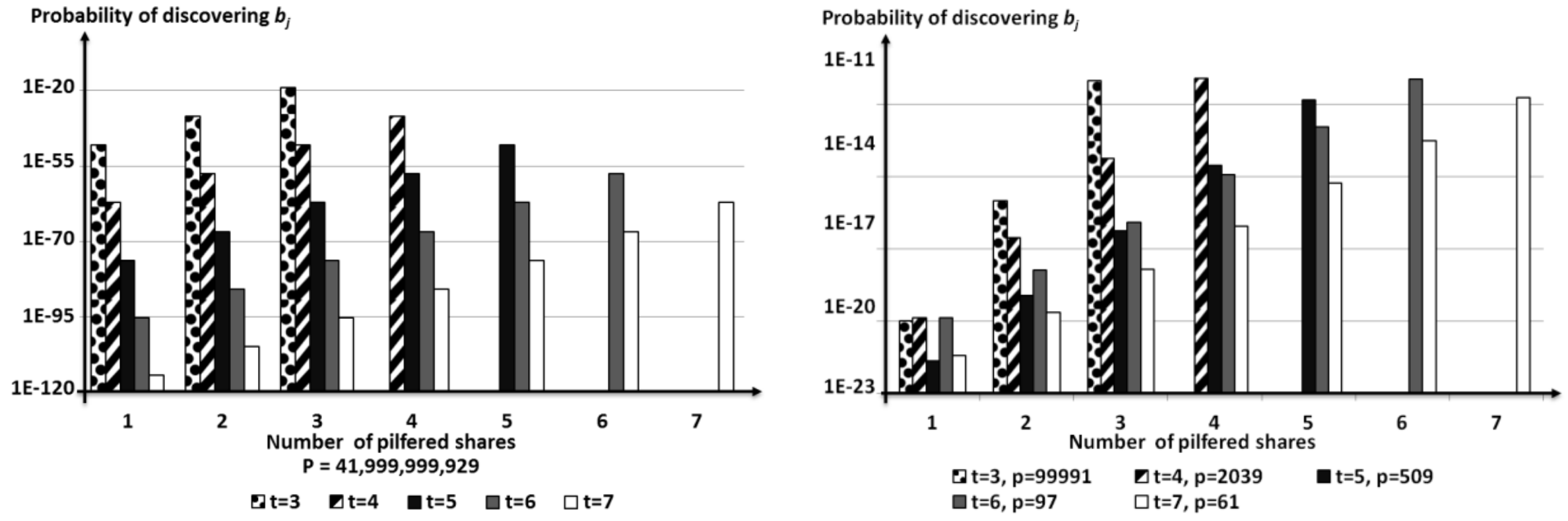

(a) Scheme-I (b) Scheme-II

*Figure 10.* Probability of decrypting a data block from its shares.

Within Scheme-II, although some data can be decrypted, if an intruder steals all data from one CSP, s/he must discover the pattern of data blocks and generate all $p^{t-1}$ combinations of data pieces stored at the $t-1$ other CSPs' by brute force. The complexity of Scheme-II's reconstructing process is $O(mt^2)$, since the $t \times t$ $C$ matrix must be computed for $m$ data pieces. Thus, with $p = 99{,}991$, $t = 3$ and $m = 100$ (11 KB of data), breaking the secret with a standard desktop computer would take more than 13 years. Thence, even with a botnet available, even partially decrypting a giga or terabyte-scale DW cannot be achieved in reasonable time.

***Reliability***

Reliability includes data availability and recovery, which are achieved by design with secret sharing, and data integrity and correctness. Our schemes can verify both the honesty of CSPs and the correctness of shares. Verification performance depends on the user-defined hash functions that define inner and outer signatures.

To test the reliability of our signatures, we generate random 32-bits unsigned integers and share them. Then, we generate errors in all shares with respect to a given pattern. Finally, we account the number of incorrect data pieces that are not detected as such. Figure 11 plots the ratio of false positives achieved with inner signature $s_in_j = \sum_{d_i \in b_j} d_i \pmod p$ and outer signature $s_out_{j,k} = e_{j,k} \bmod p_2$, where $p_2$ is a prime. If only the inner signature is used to verify data, i.e., only the honesty of CSPs is verified, the ratio ranges between 7.7E-6% and 3.2E-2%, inversely depending on $p$. However, all incorrect data pieces can be detected if data are verified by both inner and outer signatures (i.e., share correctness is also verified) and $p_2 > 61$.

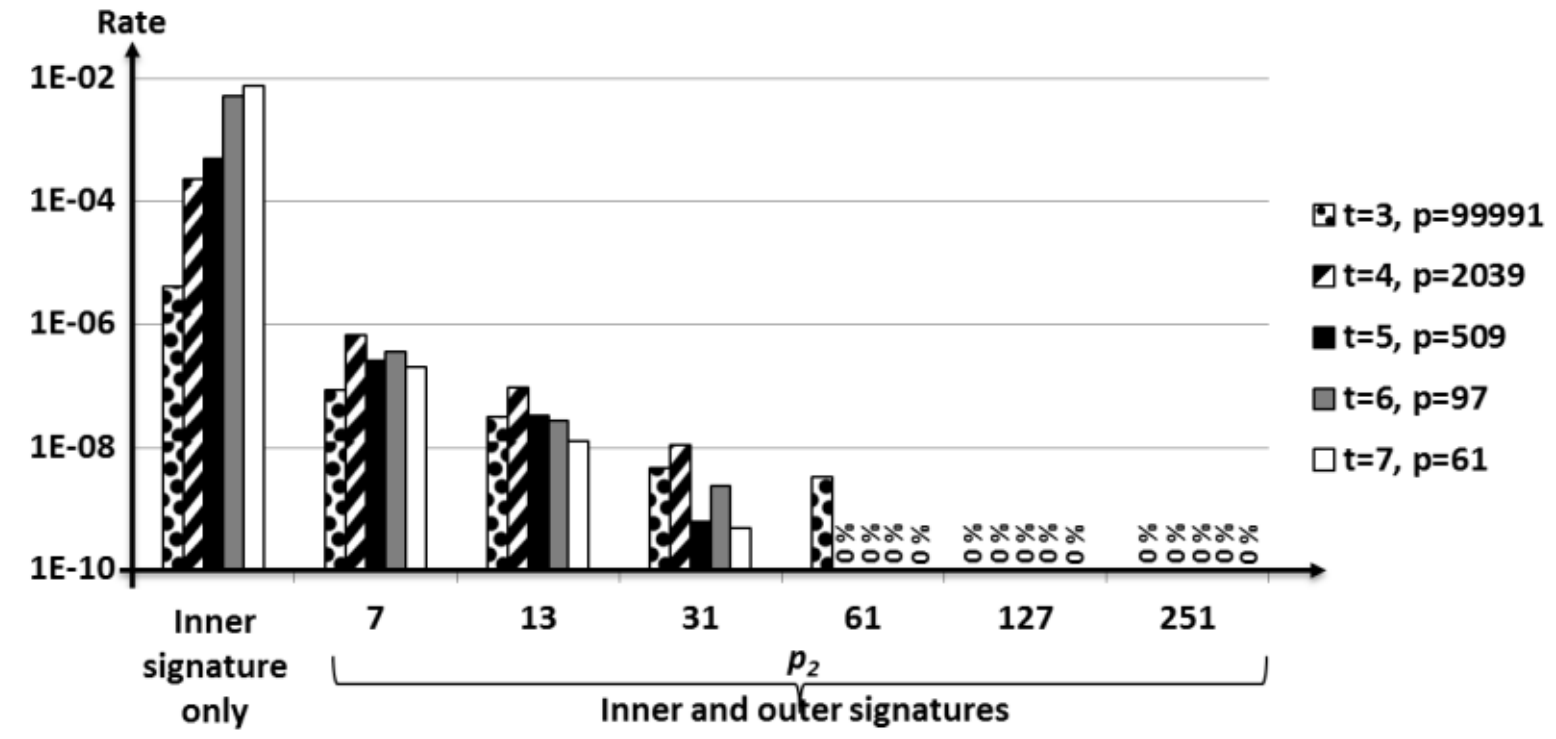

*Figure 11.*Rate of incorrect data not being detected.

Finally, note that sharing data on one node is a surjective function, i.e., two different initial values may have the same share value. However, since the reconstruction process is achieved by intersection from all nodes, sharing data is overall a bijective function. Thus, querying shares always results in a 100% hit rate.

**Cost analysis**

In this section, we provide a cost analysis of the main factors inducing costs when storing a DW in the cloud. For this sake, in addition to theoretical considerations, we run two series of experiments. In the first series of experiments, we run 100 1 GB test cases made of 32-bits unsigned integers and vary parameters $t$, $n$ and $p$. In the second series of experiments, we use the Star Schema Benchmark (SSB – O'Neil et al., 2009) and vary parameter $p$ with $t = 3$ and $n = 4$. Experiments are conducted with Bloodshed Dev-C++ 5.5.3 and MySQL 5.0.51a on a PC with an Intel(R) Core(TM) i5 2.76 GHz processor with 3 GB of RAM running Microsoft Windows 7.

***Computing cost***

The time complexity of Scheme-I's data sharing process is $O(ont)$, since $n$ $t$-variable linear equations must be computed for $o$ data blocks. Moreover, since $m = o(t-1)$, complexity can also be expressed as $O(mn)$. The time complexity of Scheme-II's data sharing process is also $O(ont)$, or $O(mnt)$ since $m = o$ here.

The time complexity of Scheme-I's data reconstruction process is $O(mt)$ or $O(ot^2)$, since the $t \times t$ $C$ matrix must be computed for $o$ data blocks and $o = \left\lceil \frac{m}{t-1} \right\rceil$. Scheme-II's is $O(mt^2)$ or $O(ot^2)$, since $C$ must be computed for $o$ data blocks and $o = m$.

For example, the execution time of sharing and reconstructing 32-bits unsigned integers with Scheme-II is plotted in Figure 12 with respect to $t$ and $n$. The execution time of both processes increase with $t$ when $t = n$. The execution time of the sharing process increases with $n$ when $t$ is fixed, whereas $n$ does not affect reconstruction. For instance, the execution time of the data sharing and reconstruction processes are about 15 seconds (throughput is 68 MB/s) and 7 seconds (throughput is 144 MB/s) when $n = t = 3$ and $p = 99991$.

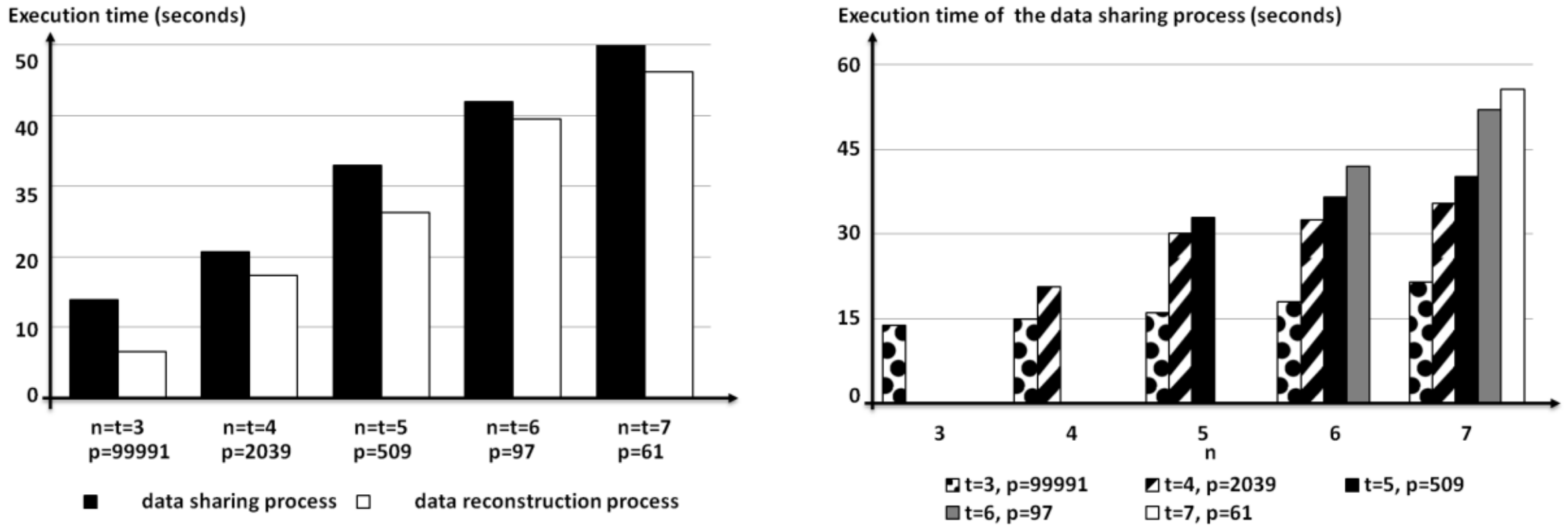

*Figure 12.*Execution time of Scheme-II.

To evaluate the performance of data analysis on shares more accurately, we measure the execution time of SSB's OLAP query workload (Table 3). Quite unexpectedly, query execution is faster on shares than on the original DW. But this is because we run plain SSB queries on the original DW and MySQL does not optimize joins natively. When executing the same query on shares, we have to split the original queries and process subqueries before joining, thus implicitly optimizing join dimensionality. Query response on shares appears reasonable, though it could be further enhanced through shared indices and materialized views.

| SSB query | Query response time (seconds) | | Query result size | |
|---|---|---|---|---|
| | Original query | Scheme-II query | Original query (bytes) | Scheme-II query (KB) |
| Q.1.1 | 8 | 10 | 12 | 7,751 |
| Q.1.2 | 4 | 4 | 11 | 176 |
| Q.1.3 | 3 | 4 | 11 | 32 |
| Q.2.1 | 80 | 25 | 6,097 | 930 |
| Q.2.2 | 80 | 7 | 1,232 | 388 |
| Q.2.3 | 80 | 4 | 1,232 | 28 |
| Q.3.1 | 78 | 49 | 2,460 | 16,343 |
| Q.3.2 | 66 | 8 | 14,400 | 864 |
| Q.3.3 | 64 | 5 | 567 | 25 |
| Q.3.4 | 64 | 4 | 72 | 5 |
| Q.4.1 | 80 | 54 | 756 | 13614 |
| Q.4.2 | 81 | 36 | 2,760 | 3531 |
| Q.4.3 | 66 | 5 | 9,936 | 11 |
| **Average** | **58** | **17** | **3,042** | **3,361** |

*Table 3.*OLAP performance with SSB.

### *Storage cost*

One advantage of our schemes is that the volume of shares nears that of original data when $n = t$, $t$ is big, and $P$ in Scheme-I and $p$ in Scheme-II are small. Shared data volume is only $on\|P\|$ in Scheme-I and $on\|p\|$ in Scheme-II, where $\|P\|$ and $\|p\|$ are sizes of $P$ and $p$, re-spectively. For example, with Scheme-II, let us consider a set $D$ of ten 32-bits unsigned integers that is shared among six CSPs, with five CSPs being sufficient to reconstruct $D$. The volume of $D$ is $10 \times 32 = 320$ bits. Let $\|p\| = 9$ bits. Then, the volume of all shares is lower than $10 \times 6 \times 9 = 540$ bits ($1.69 \times D$). The volume of each share is about $10 \times 9 = 90$ bits ($0.17 \times D$).

The volume of our shared 32-bit unsigned integer dataset using Scheme-II is plotted in Figure 13. The volume of each share varies with respect to $n$, $t$ and $p$. The volume of all shares ranges between $\frac{n}{t-1}$ and $n$ times the volume of $D$. For example, it is 1.89 GB ($\frac{3}{3-1} = 1.5 \leq 1.89 \leq 3$) when the volume of original data is 1 GB, $n = t = 3$ and $p = 99991$.

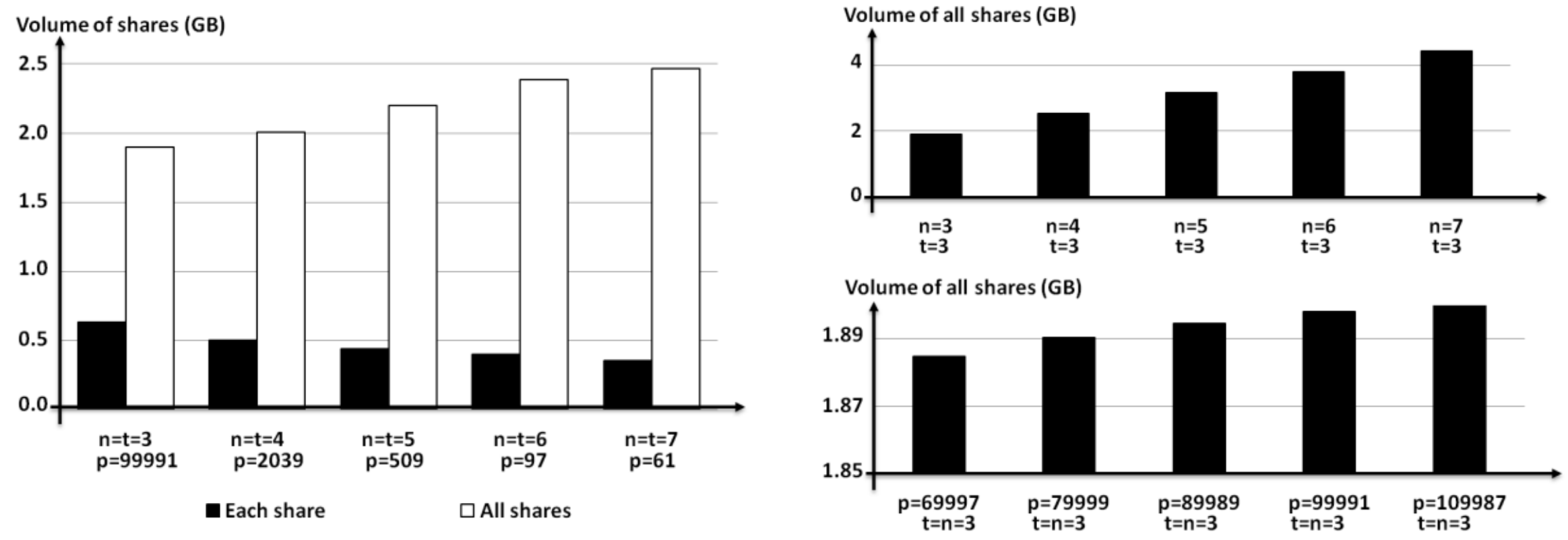

*Figure 13.* Shared data volume.

Finally, Table 4 shows the volume of SSB's DW, once shared with Scheme-II. The volume of each share is still smaller than that of the original DW (about 46% smaller). Therefore, this guarantees any shared DW can be stored and queried if the original DW can. If data volume is very large, a higher-performance DBMS can be envisaged, e.g., a parallel DBMS or low-level distributed storage. Although, the volume of all shares is greater than that of original data (about 185% greater), it is smaller than twice that of original data. Shared volume may be reduced to about 139% of original data if the data availability constraint is relaxed, though.

| **Table** | **Original SSB** | **1st share** | | **2nd share** | | **3th share** | | **4th share** | | **All shares** | |
|---|---|---|---|---|---|---|---|---|---|---|---|
| | **Volume (KB)** | **Volume (KB)** | **%[(1)]** | **Volume (KB)** | **%[(1)]** | **Volume (KB)** | **%[(1)]** | **Volume (KB)** | **%[(1)]** | **Volume (KB)** | **%[(1)]** |
| Customer | 3,167 | 2,550 | 80.52 | 2,550 | 80.52 | 2,550 | 80.52 | 2,550 | 80.52 | 10,200 | 322.07 |
| Date | 218 | 205 | 94.04 | 205 | 94.04 | 205 | 94.04 | 205 | 94.04 | 820 | 376.15 |
| Part | 18,798 | 14,940 | 79.48 | 14,940 | 79.48 | 14,940 | 79.48 | 14,940 | 79.48 | 59,760 | 317.91 |
| Supplier | 965 | 768 | 79.59 | 758 | 78.55 | 758 | 78.55 | 758 | 78.55 | 3,042 | 315.23 |
| Lineorder | 822,929 | 373,610 | 45.40 | 373,610 | 45.40 | 373,610 | 45.40 | 373,610 | 45.40 | 1,494,440 | 181.60 |
| **All tables** | **846,077** | **392,073** | **46.34** | **392,063** | **46.34** | **392,063** | **46.34** | **392,063** | **46.34** | **1,568,262** | **185.36** |

[(1)]Percentage of original data volume.

*Table 4.* SSB shared data warehouse volume achieved with Scheme-II.

***Data transfer cost***

In our context, data transfer cost only relates to the size of query results, since uploads are generally free of charge at major CSPs'. SSB query result size is shown in Table 3. Since many queries do not need to decrypt data, only some parts of the shared DW are transferred when executing SSB's workload. Thus, the transferred data volume is greater than the volume of each query result, but lower than that of all shares. For example, query Q.1.3 ran on shares outputs 32 KB, which is greater the actual result size (11 bytes) but much lower than the volume of all shares (1.5 GB), thus incurring reasonable transfer costs.

## Comparison of Scheme-II to existing related approaches

In this section, we compare Scheme-II to the TWH approaches presented in our state of the art, with respect to security and cost. Table 5 synthesizes the features and costs of all approaches, which we discuss below.

| Features and costs | Thompson et al. (2009) | Wang et al. (2011) | Hadavi et al. (2012) | Scheme-II |
|---|---|---|---|---|
| Privacy | Yes | Yes | Yes | Yes |
| Data availability | Yes | Yes | Yes | Yes |
| Data integrity | | | | |
| - Inner code verifying | Yes | Yes | Yes | Yes |
| - Outer code verifying | No | No | No | Yes |
| Target | Databases | Databases | Databases | Data warehouses |
| Data types | Positive integers | Positive integers | Positive integers | Integers, Reals, Characters, Strings, Dates, Booleans |
| Shared data access | | | | |
| - Data updates | Yes | Yes | Yes | Yes |
| - Exact match queries | No | Yes | Yes | Yes |
| - Range queries | No | Yes | Yes | No |
| - Aggregation functions | Yes | No | Yes | Yes |
| - OLAP queries | No | No | No | Yes |
| Costs | | | | |
| - Data storagew.r.t. original data volume | $\geq 2n$ + 1 (hash tree) | $\geq n/t$ + $n/t$ (B++ tree index) + signatures | $\geq n$ + 1 (B++ tree index) | $\geq n/(t-1)$ + signatures |
| - Sharing process execution time | $O(mnt)$ | $O(\max(m \log m, mn))$ | $O(mnt)$ | $O(mnt)$ |
| - Reconstructing process execution time | $O(mt^2)$ | $O(mt)$ | $O(mt^2)$ | $O(mt^2)$ |

*Table 5.* Comparison of database sharing approaches

### *Security*

All approaches handle data security, availability and integrity issues, but only ours verifies both the correctness of shares and the honesty of CSPs by outer and inner code verification, respectively.

### *Computing cost*

The time complexity of Scheme-II's sharing process is equal to TH's and better than W's because $m$ is normally much bigger than $n$ and $t$. Thence, $m \log m > mnt > mn$. However, in the context of DWs, where updates are performed off-line, update performance is not as critical as in transactional databases.

The time complexity of Scheme-II's reconstructing process is again equal to TH's and better than W's. Since data decryption is part of query response time, it is critical in a DW context. However, shared data access is also part of query response time. In this regard, our approach is faster than TWH, because we can directly query shared tables, whereas TWH must perform ad-hoc queries, aggregate and reconstruct data to achieve the same result. For instance, W cannot perform any aggregation operation on shares. Thence, many shares are transferred back to the user for aggregation.

### *Storage cost*

In addition to the storage estimations provided in Table 5, let us illustrate the storage gain achieved with our approach through an example. Let $n = 4$, $t = 3$ and $V$ be the original data volume. Let us also assume that each share is not bigger than the original data it encrypts. For simplicity, let us finally disregard the volume of signatures that depends on user-defined parameters in all approaches. The result is shown in Table 6, with column *Improvement* displaying the storage gain achieved by Scheme-II over TWH.

| Approach | Shared data volume | Scope | Improvement |
|---|---|---|---|
| Scheme-II | $\frac{n}{(t-1)}V = \frac{4}{(3-1)}V = 2V$ | Shares | 0% |
| T | $2nV = 2 \times 4 \times V = 8V$ | Shares | 300% |
| W | $\frac{n}{t}V + \frac{n}{t}V = \frac{2n}{t}V = \frac{2 \times 4}{3}V = 2.67V$ | Shares and indices | 33% |
| H | $nV + V = (n+1)V = (4+1) \times V = 5V$ | Shares and indices | 150% |

*Table 6.* Comparison shared data volume in Scheme-II and TWH.

### *Data transfer cost*

Data transfer cost directly relates to the size of shares when loading data, and to the size of query results when accessing the shared database. Since all approaches allow different operations and vary in shared data volume, it is difficult to compare data transfer costs by proof. However, data transfer cost in our approach is cheaper in the sharing phase because the size of each encrypted data piece is $1/(t-1)$ smaller than that of TWH. Moreover, by creating shared data cubes, we allow straight computations on shares, and thus only target results are transferred to the user, i.e., with no additional data to decrypt at the user's.

## CONCLUSION

In this paper, we propose an original approach to share a DW in the cloud that simultaneously supports data privacy, availability, integrity and OLAP querying. Our approach is constituted of two schemes. Scheme-I exploits block cryptography and secret sharing to protect data and guarantee data privacy and availability. Moreover, Scheme-I ensures data correctness by utilizing homomorphic and hash functions as signatures. Scheme-II builds upon Scheme-I to allow sharing and querying cloud DWs. It allows analyzing data over shares without decrypting all data first. Our security and performance analysis shows that our schemes are more robust and cheaper than similar existing techniques when storing and querying data.

Future research shall run along two axes. First, we plan to further assess the cost of our solution in the cloud pay-as-you-go paradigm. Sharing data indeed implies increasing the initial data volume, and thus storage cost, as well as duplicating computing costs over CSPs. However, it also guarantees data availability. Hence, we plan to run monetary cost evaluations against classical data replication schemes. It would also be very interesting to balance the cost of our solution against the cost of risking data loss or theft. Moreover, parameter assignment affects the security of our schemes. Notably, to enforce security, big values should be assigned to primes $P$, $p$ and number of CSPs needed to decrypt data $t$. In contrast, small values should be assigned to $P$, $p$, $n$ and $t$ to reduce execution time and data volume. Thus, a suitable tradeoff must be investigated.

Second, although we provide in this paper a raw framework for OLAPing shared data, more research is required to implement all operations needed in OLAP analyses, as well as incremental updates. We notably plan to reuse the strategies of Wang et al. (2011) and Hadavi et al. (2012) to achieve range and match queries, e.g., by implementing shared B+ tree indices.